\documentclass[12pt,letterpaper]{article}
\usepackage{epsfig,rotating,setspace,latexsym,amsmath,epsf,amssymb,bm, color}
\usepackage{cite}

\title{Capacity of a Class of Diamond Channels\thanks{This work was
supported by NSF Grants CCF $04$-$47613$, CCF
$05$-$14846$, CNS $07$-$16311$ and CCF $07$-$29127$.}}

\author{Wei Kang \qquad Sennur Ulukus \\
\normalsize Department of Electrical and Computer Engineering\\
\normalsize University of Maryland, College Park, MD 20742 \\
\normalsize {\it wkang@umd.edu} \qquad {\it ulukus@umd.edu}}

\newtheorem{Theo}{Theorem}

\begin{document}
%\date{}
\maketitle 

\begin{abstract}
We study a special class of diamond channels which was introduced by Schein in 2001. In this special class, each diamond channel consists of a transmitter, a noisy relay, a noiseless relay and a receiver. We prove the capacity of this class of diamond channels by providing an achievable scheme and a converse. The capacity we show is strictly smaller than the cut-set bound. Our result also shows the optimality of a combination of decode-and-forward (DAF) and compress-and-forward (CAF) at the noisy relay node. This is the first example where a combination of DAF and CAF is shown to be capacity achieving. Finally, we note that there exists a duality between this diamond channel coding problem and the Kaspi-Berger source coding problem.
\end{abstract}

\newpage
\section{Problem Statement and the Result}
\begin{figure}[h]
\centering
\input{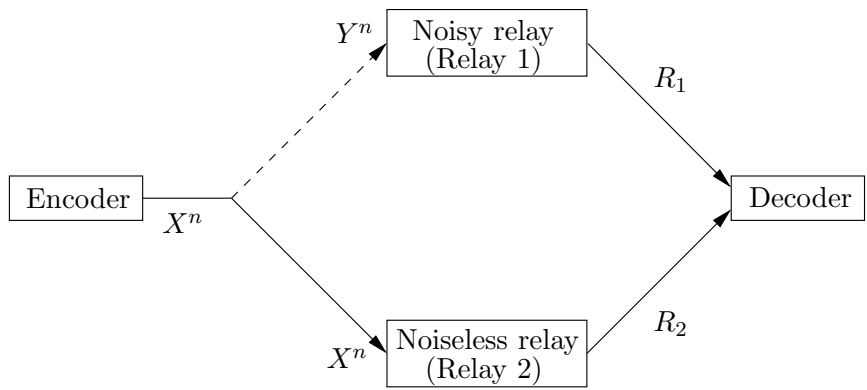}
\caption{The diamond channel.}
\label{DChannel}
\end{figure}

The diamond channel was first introduced by Schein in 2001 \cite{Schein:2001}. The diamond channel consists of one transmitter, two relays and a receiver, where the transmitter and the two relays form a broadcast channel as the first stage and the two relays and the receiver form a multiple access channel as the second stage. 
The capacity of the diamond channel in its most general form is open. Schein explored several special cases of the diamond channel, one of which \cite[Section 3.5]{Schein:2001} is specified as follows (see Figure \ref{DChannel}). The multiple access channel consists of two orthogonal links with rate constraints $R_1$ and $R_2$, respectively. The broadcast channel contains a noisy branch and a noiseless branch, i.e.,~with input $X$ and two outputs $X$ and $Y$. We refer to the relay node receiving $Y$ as the noisy relay and the relay node receiving $X$ as the noiseless relay. Schein provided two achievable schemes for this class of diamond channels. In this paper, we will prove the capacity of this special class of diamond channels.

The formal definition of the problem is as follows.
Consider a channel with input alphabet $\mathcal{X}$ and output alphabet $\mathcal{Y}$, which is characterized by the transition probability $p(y|x)$. Assume an $n$-length block code consisting of $(f,g,h,\varphi)$ where
\begin{align}
f:&\{1,2,\dots, M\}\mapsto\mathcal{X}^n\\
g:& \mathcal{Y}^n\mapsto \{1,2,\dots, |g|\}\\
h:& \{1,2,\dots, M\}\mapsto\{1,2,\dots,|h|\}\\
\varphi:& \{1,2,\dots, |g|\}\times\{1,2,\dots,|h|\}\mapsto\{1,2,\dots,M\}
\end{align}
Here $f$ denotes the encoding function at the transmitter, $g$ and $h$ denote the processing functions at the noisy and noiseless relays, respectively, and $\varphi$ denotes the decoding function at the receiver.

The encoder sends $x^n=f(m)$ into the channel, where $m\in\{1,2,\dots,M\}$. The decoder reconstructs $\hat{m}=\varphi(g(Y^n), h(m))$. The average probability of error is defined as
\begin{equation}
P_e\triangleq\frac{1}{M}\sum_{m=1}^M Pr(\hat{m}\ne m|m \text{ is sent})
\end{equation}
The rate triple $(R,R_1,R_2)$ is achievable if for every $0<\epsilon<1$, $\eta>0$ and every sufficiently large $n$, there exists an $n$-length block code $(f,g,h,\varphi)$, such that $P_e\le \epsilon$ and
\begin{align}
\frac{1}{n}\ln M&\ge R-\eta\\
\frac{1}{n}\ln |g|&\le R_1+\eta\\
\frac{1}{n}\ln |h|&\le R_2+\eta 
\end{align}

The following theorem characterizes the capacity of the class of diamond channels considered in this paper. 

\begin{Theo}\label{maintheo}
The rate triple $(R, R_1,R_2)$ is achievable in the above channel if and only if the following conditions are satisfied
\begin{align}
R&\le I(U;Y)+H(X|U)\label{dc1}\\
R_1&\ge I(Z;Y|U,X)\label{dc2}\\
R_2&\ge H(X|Z,U)\label{dc3}\\
R_1+R_2&\ge R+I(Y;Z|X,U)\label{dc4}
\end{align}
for some joint distribution 
\begin{equation}
p(u,z,x,y)=p(u,x)p(y|x)p(z|u,y)
\end{equation}
with cardinalities of alphabets satisfying
\begin{align}
|\mathcal{U}|&\le|\mathcal{X}|+4\label{dcb1}\\
|\mathcal{Z}|&\le|\mathcal{U}||\mathcal{Y}|+3\le|\mathcal{X}||\mathcal{Y}|+4|\mathcal{X}|+3\label{dcb2}
\end{align}
\end{Theo}

\section{The Achievability}
Assume a given joint distribution
\begin{equation}
p(u,z,x,y)=p(u,x)p(y|x)p(z|u,y)
\end{equation}
and consider that the information theoretic quantities on the right hand sides of (\ref{dc1}), (\ref{dc2}), (\ref{dc3}) and (\ref{dc4}) are evaluated with this fixed joint probability distribution.
%We note that it suffices to prove the  boundary case where $R=I(U;Y)+H(X|U)$. 
%Consider a message $W$ with size of alphabet $M$ such that 
%\begin{equation}
%\frac{1}{n}\ln M= I(U;Y)+H(X|U)-\eta
%\end{equation}

\begin{figure}[t]
\centering
\input{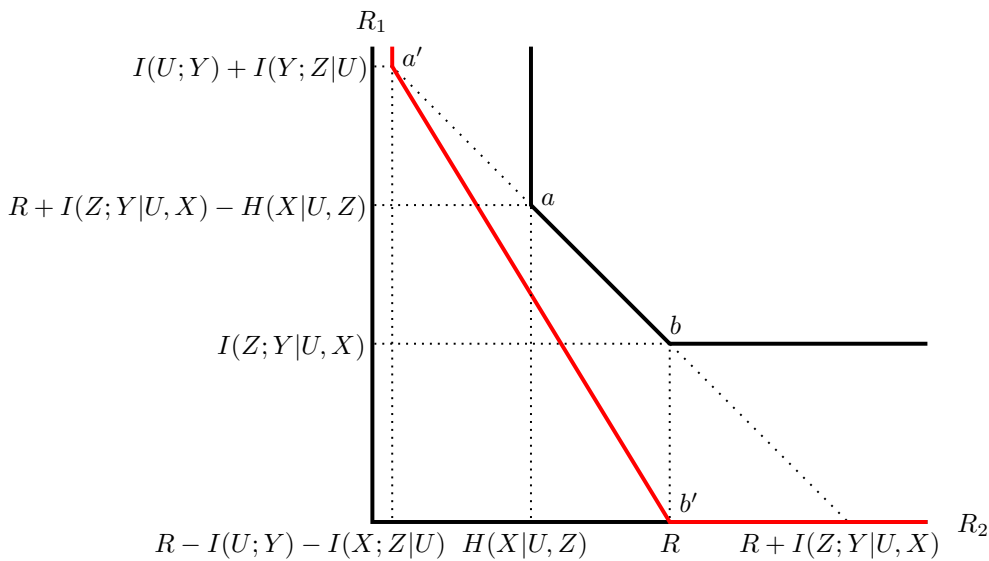}
\caption{Rate region of $(R_1,R_2)$ when $H(X|U,Z)\le R\le I(U;Y)+I(X;Z|U)$.}
\label{Rregion}
\end{figure}

Consider a message $W$ with rate $R$. If $R\le H(X|Z,U)$, reliable transmission can be achieved by letting $g(Y^n)=\phi$ (constant) and $h(W)=W$, i.e.,~by sending the message through the noiseless relay.  Thus, we will only consider the case where
\begin{equation}
H(X|Z,U)<R\le I(U;Y)+H(X|U)
\end{equation}
We will show that the message can be reliably transmitted with a pair of functions $(g,h)$  such that $(\frac{1}{n}\ln|g|,\frac{1}{n}\ln|h|)$ lies in the inverse pentagon\footnote{By ``inverse pentagon'' with corner points $a$ and $b$, we mean the region in the $(R_1,R_2)$ space that is to the ``north-east'' of line segment $[a,b]$. More specifically, this is the region described by inequalities in (\ref{dc2}), (\ref{dc3}) and (\ref{dc4}).} with corners $a$ and $b$ in Figure \ref{Rregion}. However, we instead prove  reliable transmission with $(\frac{1}{n}\ln|g|,\frac{1}{n}\ln|h|)$ lying in the inverse pentagon with corners $a'$ and $b'$, which contains the inverse pentagon with corners $a$ and $b$ and thus imposes a stronger condition to prove. %(similar argument can be found in \cite[Section V, Case 2]{Berger:1989}). 
It is straightforward to have reliable transmission with the rate pair at point $b'$ by letting $g(Y^n)=\phi$ (constant) and $h(W)=W$. Thus, it remains to prove that  reliable transmission is possible  with the rate pair at point $a'$,
%if reliable transmission is possible with rate pair at the point $a$, while the rate pair on the red line can be achieved by time sharing, we will prove that the message can be reliable transmitted with the rate pair in a larger pentagon with corners $a$ and $d$.  Now we will show the reliable transmission with the rate pair $(R_1, R_2)$ at the point $a$, 
i.e.,
\begin{align} 
R_1&=I(U;Y)+I(Y;Z|U)\\
R_2&=R-I(U;Y)-I(X;Z|U)
\end{align}
%such that for given rate pair $(R_1, R_2)$,
%\begin{align}
%R_1&\ge I(Z;Y|U)\\
%R_2&\ge H(X|Z,U)
%\end{align}
%We define
%\begin{align}
%R_1'&\triangleq R_1-I(Z;Y|U)\\
%R_2'&\triangleq R_2-H(X|Z,U)
%\end{align}
%Consider a message $W$ with the rate 
%\begin{equation}
%R\triangleq \frac{1}{n}\ln|W|=\min(R_1+R_2-I(Y;Z|X,U),I(U;Y)+H(X|U))
%\end{equation}

Let us assume that the message $W$ is decomposed as $W=(W_a,W_b,W_c)$. For a positive number $\epsilon $, let us define 
\begin{align}
M_a\triangleq|W_a|&=\exp(n(I(U;Y)-3 \epsilon))\\
M_b\triangleq|W_b|&=\frac{M}{M_aM_c}=\exp(\ln M-n(I(U;Y)+I(X;Z|U)+6\epsilon))\\
M_c\triangleq|W_c|&=\exp(n(I(X;Z|U)-3 \epsilon))
\end{align}

\emph{Random codebook generation}: We use a superpostion code structure. The size of the inner code is $M_a$. For each inner codeword, we independently generate $M_b$ outer codes. The size of each outer code is $M_c$.
\begin{itemize}
\item Independently generate $M_a$ sequences, $u^n(1),u^n(2),\dots, u^n(M_a)$, according to $\prod_{i=1}^n p(u_i)$ where $p(u_i)=p(u)$, for $i=1,2,\dots, n$. 
\item For $u^n(j)$, $j=1,2,\dots, M_a$, independently generate $M_b$ codebooks, $\mathcal{C}(j,1),\mathcal{C}(j,2), \dots$, $ \mathcal{C}(j,M_b)$. 
\item In the  codebook $\mathcal{C}(j,k)$, $j=1,2,\dots,M_a$, $k=1,2,\dots, M_b$, independently generate $M_c$ codewords $x^n(j,k,1), x^n(j,k,2),\dots, x^n(j,k,M_c)$ according to $\prod_{i=1}^n p(x_i|U_i=u_{i}(j))$, where $p(x_i|U=u_{i}(j))=p(x|u)$, for $i=1,2,\dots,n$, $j=1,2,\dots,M_a$, $k=1,2,\dots, M_b$.
\end{itemize}
There will be no overlapping codebooks with high probability when $n$ is sufficiently large, because
\begin{equation}
\frac{1}{n}\ln M_bM_c<H(X|U)
\end{equation}

\emph{Encoding at the transmitter}: Let $W=(W_a,W_b,W_c)$ be the message. We send codeword $X^n=f(W_a,W_b,W_c)\triangleq x^n(W_a,W_b,W_c)$ into the channel.

\emph{Processing at the noisy relay}: First, after having received $Y^n$, seek 
\begin{equation}
\hat{U}^n=u^n(\hat{W}_a)\in\{u^n(1),u^n(2),\dots, u^n(M_a)\}
\end{equation} 
such that   
\begin{equation}
(\hat{U}^n,Y^n)\in \mathcal{T}_{[UY] }^n
\end{equation}
where the definition of strong typical set can be found in \cite[Section 1.2]{Csiszar:1981}. If there is not any such $\hat{U}^n$, then let $\hat{U}^n$ be an arbitrary sequence in $\{u^n(1),u^n(2),\dots, u^n(M_a)\}$. 
%Secondly, independently generate $L$ sequences, $\{z^n(1,U^n), z^n(2,U^n),\dots, z^n(L,U^n)\}$, each of which is uniformly distributed in the conditional strong typical set $\mathcal{T}_{[Z|U] }^n(U^n)$ (the definition of strong typical set can be found in \cite[Section 1.2]{Csiszar:1981}). Assume $\frac{1}{n}\ln L= I(Y;Z|U)+ $, then there exists a function $f'$, where $Z^n=f'(Y^n,U^n)$, when $n$ is sufficiently large, such that 
%\begin{equation}
%Pr((f'(Y^n,U^n), Y^n)\in \mathcal{T}_{[YZ|U] }^n(U^n) |U^n\in\mathcal{T}_{[U] }^n)\ge 1- 
%\end{equation}
%if $Y^n$ is generated according to $\prod_{i=1}^n p(y_i|u_i(W_a)))$, and $ \rightarrow0$ when $ \rightarrow0$.  
Secondly, construct a conditional rate distortion code according to $\prod_{i=1}^np(z_i,y_i|\hat{u}_i)$ with encoding function $g'(Y^n,\hat{U}^n)$ and $|g'|=L=\exp(n(I(Y;Z|U)+\tau))$.
Finally send $\hat{U}^n$ and $Z^n\triangleq g'(Y^n,\hat{U}^n)$ to the destination, i.e.,
\begin{equation}
g(Y^n)=(\hat{U}^n,Z^n)
\end{equation}
where
\begin{equation}
|g|=M_a\times L\le\exp(n(I(U;Y)+I(X;Z|U)+ \tau-3\epsilon))
\end{equation}

\emph{Processing at the noiseless relay}: Let $h(f(W_a,W_c,W_b))=W_b$ where\begin{equation}
|h|=M_b=\exp(\ln M-n(I(U;Y)+I(X;Z|U)+6\epsilon))
\end{equation}

\emph{Decoding}: Decoder collects $(\hat{U}^n,Z^n)$ from the noisy relay and $W_b$ from the noiseless relay. The decoder seeks  a codeword $x^n(W_a,W_b,i)$ from the codebook $\mathcal{C}(W_a,W_b)$ such that
\begin{equation}
(x^n(\hat{W}_a,W_b,i), Z^n)\in \mathcal{T}_{[XZ|U] }^n (\hat{U}^n)
\end{equation}

\emph{Probability of error}: The error occurs when $(\hat{U},\hat{X})\ne(U,X)$.
The average probability of error can be decomposed into
\begin{equation}
Pr(E)\le Pr(E_1\cup E_2\cup E_3)=Pr(E_1)+Pr(E_2 \cap E_1^c)+Pr(E_3 \cap E_1^c\cap E_2^c)
\end{equation}
where
\begin{align}
E_1&\triangleq (U^n,X^n,Y^n,Z^n)\notin \mathcal{T}_{[UXYZ] }^n\\
E_2&\triangleq \bigcup_{\bar{u}^n\ne U^n, \bar{u}^n\in\{u^n(1),u^n(2),\dots,u^n(M_a)\}}(\bar{u}^n,Y^n)\in \mathcal{T}_{[UY] }^n\\
E_3&\triangleq \bigcup_{\bar{x}^n\ne X^n, \bar{x}^n\in\mathcal{C}(W_a,W_b)}(\bar{x}^n,Z^n)\in \mathcal{T}_{[XZ|U] }^n(U^n)
\end{align}
We note that
\begin{align}
Pr(E_1)&\le Pr(U^n\notin\mathcal{T}_{[U] }^n)+Pr((Y^n,Z^n)\notin\mathcal{T}_{[YZ|U] }^n(U^n))+Pr(X^n\notin\mathcal{T}_{[X|YZU] }^n(Y^n,Z^n,U^n))
\end{align}
where 
\begin{itemize}
\item $U^n$ is generated in an i.i.d. fashion with probability $p(u)$. Thus, when $n$ is sufficiently large, we have
\begin{equation}
Pr(U^n\notin\mathcal{T}_{[U] }^n)\le  \epsilon
\end{equation}
\item $Z^n$ is a conditional rate distortion code for $Y^n$ conditioned on $U^n$. Thus, when $n$ is sufficiently large, $L=\exp(nI(Y;Z|U)+\tau)$,  and $U^n\in\mathcal{T}_{[U] }^n$, we have
\begin{equation}
Pr((Y^n,Z^n)\notin\mathcal{T}_{[YZ|U] }^n(U^n))\le  \epsilon
\end{equation}
\item $X^n$ can be viewed as being generated according to an i.i.d. conditional probability $p(x|u,y)$ with respect to $(U^n,Y^n)$.
 Thus, when $n$ is sufficiently large and $(Y^n,Z^n,U^n) \in \mathcal{T}_{[YZU]}^n$,
\begin{equation}
Pr(X^n\notin\mathcal{T}_{[X|YZU] }^n(Y^n,Z^n,U^n))\le  \epsilon
\end{equation}
\end{itemize}
From the above calculation, we have
\begin{equation}
Pr(E_1)=Pr((U^n,X^n,Y^n,Z^n)\notin \mathcal{T}_{[UXZ] }^n)\le 3 \epsilon
\end{equation}
For the second error event, we note that $M_a=\exp(n(I(U;Y)-3 \epsilon)$ and
\begin{align}
Pr(E_2\cap E_1^c)&=Pr\left(\bigcup_{\bar{u}^n\ne U^n, \bar{u}^n\in\{u^n(1),u^n(2),\dots,u^n(M_a)\}}(\bar{u}^n,Y^n)\in \mathcal{T}_{[UY] }^n|(Y^n)\in \mathcal{T}_{[Y] }^n\right)\nonumber\\
&\le\sum_{i=1}^{M_a}Pr((u^n(i),Y^n)\in \mathcal{T}_{[UY] }^n|Y^n\in \mathcal{T}_{[Y] }^n)\nonumber\\
&\le M_a Pr(u^n(i)\in\mathcal{T}_{[U|Y] }^n(Y^n))\nonumber\\
&\le M_a \exp(-nH(U)+n \epsilon)\exp(nH(U|Y)+n \epsilon)\nonumber\\
&= \exp(-n \epsilon)\nonumber\\
&\le \epsilon \qquad  
\end{align}
for  sufficiently large $n$. We note that $M_c=\exp(n(I(X;Z|U)-3 \epsilon)$, then
\begin{align}
Pr(E_3\cap E_1^c)&=Pr\left(\bigcup_{\bar{x}^n\ne X^n, \bar{x}^n\in\mathcal{C}(W_a,W_b)}(\bar{x}^n,Z^n)\in \mathcal{T}_{[XZ|U] }^n(U^n)|(Z^n,U)\in \mathcal{T}_{[ZU] }^n\right)\nonumber\\
&\le\sum_{i=1}^{M_c}Pr((x(M_a,M_b,i),Z^n)\in \mathcal{T}_{[XZ|U] }^n(U^n)|(Z^n,U^n)\in \mathcal{T}_{[ZU] }^n)\nonumber\\
&\le M_c Pr(x(M_a,M_b,i)\in\mathcal{T}_{[X|ZU] }^n(Y^n))\nonumber\\
&\le M_c \exp(-nH(X|U)+n \epsilon)\exp(nH(X|Z,U)+n \epsilon)\nonumber\\
&= \exp(-n \epsilon)\nonumber\\
&\le\epsilon  \qquad 
\end{align}
for sufficiently large $n$.
Thus, the average probability error is upper bounded as
\begin{equation}
Pr(E)\le 3\epsilon + \epsilon+ \epsilon=5\epsilon 
\end{equation}
which goes to zero when $n$ goes to infinity.

\section{The Converse}
Define $Z_i\triangleq g$ and $U_i\triangleq (Y^{i-1},X_{i+1}^n)$. We note that
\begin{equation}
p(u_i,x_i,y_i,z_i)=p(u_i,x_i)p(y_i|x_i)p(z_i|y_i,u_i)
\end{equation}
We have 
\begin{align}
\ln M&=H(X^n)\nonumber\\
&=\sum_{i=1}^nH(X_i|X_{i+1}^n)\nonumber\\
&\le \sum_{i=1}^nI(Y^{i-1};Y_i)+H(X_i|X_{i+1}^n)\nonumber\\
&=\sum_{i=1}^nI(Y^{i-1},X_{i+1}^n;Y_i)-I(X_{i+1}^n;Y_i|Y^{i-1})+H(X_i|Y^{i-1},X_{i+1}^n)+I(Y^{i-1};X_i|X_{i+1}^n)\nonumber\\
&\overset{\ref{c1}}{=}\sum_{i=1}^nI(Y^{i-1},X_{i+1}^n;Y_i)+H(X_i|Y^{i-1},X_{i+1}^n)\nonumber\\
&=\sum_{i=1}^nI(U_i;Y_i)+H(X_i|U_i)\label{rate}
\end{align}
where
\begin{enumerate}
\item Because of the following equality \cite[Lemma 7]{Csiszar:1978}
\label{c1}
\begin{equation}
\sum_{i=1}^nI(X_{i+1}^n;Y_i|Y^{i-1})=\sum_{i=1}^nI(Y^{i-1};X_i|X_{i+1}^n)
\end{equation} 
\end{enumerate}
We have
\begin{align}
\ln|g|&\ge H(g)\nonumber\\
&\ge H(g|h)\nonumber\\
&\ge H(g|h)-H(g|h,  Y^n)\nonumber\\
&=I(g;Y^n|h)\nonumber\\
&=\sum_{i=1}^n I(g;Y_i|h,Y^{i-1})\nonumber\\
&=\sum_{i=1}^n I(g,X_{i+1}^n;Y_i|h,Y^{i-1})-I(X_{i+1}^n;Y_i|g,h,Y^{i-1})\nonumber\\
&\overset{\ref{c21}}{=}\sum_{i=1}^n I(g,X_{i+1}^n;Y_i|h,Y^{i-1})-I(Y^{i-1};X_i|g,h,X_{i+1}^n)\nonumber\\
&\ge \sum_{i=1}^n I(g,X_{i+1}^n;Y_i|h,Y^{i-1})-H(X_i|g,h,X_{i+1}^n)\nonumber\\
&=-H(X^n|g,h)+\sum_{i=1}^n I(g,X_{i+1}^n;Y_i|h,Y^{i-1})\nonumber\\
&\overset{\ref{c22}}{\ge}\sum_{i=1}^n I(g,X_{i+1}^n;Y_i|h,Y^{i-1})- \epsilon\nonumber\\
&\ge\sum_{i=1}^n I(g;Y_i|h,Y^{i-1},X_{i+1}^n)- \epsilon\nonumber\\
%\ge&\sum_{i=1}^n I(f;Y_i|g,Y^{i-1},X_{i+1}^n, X_i)- \nonumber\\
&\overset{\ref{c23}}{\ge}\sum_{i=1}^n I(g;Y_i|h,Y^{i-1},X_{i+1}^n, X_i)- \epsilon\nonumber\\
&\overset{\ref{c24}}{=}\sum_{i=1}^n I(g;Y_i|Y^{i-1},X_{i+1}^n, X_i)- \epsilon\nonumber\\
&=\sum_{i=1}^n I(Z_i;Y_i|U_i, X_i)- \epsilon\label{rateg}
\end{align}
where
\begin{enumerate}
\item Because of the following equality \cite[Lemma 7]{Csiszar:1978}
\label{c21}
\begin{equation}
\sum_{i=1}^nI(X_{i+1}^n;Y_i|g,h,Y^{i-1})=\sum_{i=1}^nI(Y^{i-1};X_i|g,h,X_{i+1}^n)
\end{equation}
\item Due to Fano's inequality. \label{c22}
\item $g$ is a deterministic function of $Y^n$. Due to the memoryless property, we have\label{c23}
\begin{equation}
H(g|Y_i,h,Y^{i-1},X_{i+1}^n, X_i)=H(g|Y_i,h,Y^{i-1},X_{i+1}^n)
\end{equation}
\item $g$ is a deterministic function of $Y^n$ and $h$ is a deterministic function of $X^n$. Due to the memoryless property, we have\label{c24}
\begin{align}
H(g|h,Y^{i-1},X_{i+1}^n, X_i)&=H(g|Y^{i-1},X_{i+1}^n, X_i)\\
H(g|h,Y^{i-1},X_{i+1}^n, X_i, Y_i)&=H(g|Y^{i-1},X_{i+1}^n, X_i, Y_i)
\end{align}
\end{enumerate}
We have
\begin{align}
\ln|h|&\ge H(h|g)\nonumber\\
&\ge I(h;X^n|g)\nonumber\\
&=H(X^n|g)-H(X^n|g,h)\nonumber\\
&\overset{\ref{c31}}\ge H(X^n|g)-n \epsilon\nonumber\\
&=\sum_{i=1}^n H(X_i|X_{i+1}^n,g)-\epsilon\nonumber\\
&\ge\sum_{i=1}^n H(X_i|Y^{i-1},X_{i+1}^n,g)-\epsilon\nonumber\\
&=\sum_{i=1}^n H(X_i|U_i,Z_i)-\epsilon\label{rateh}
%&= H(X^n)-H(X^n)+H(X^n|f)-n \epsilon\nonumber\\
%&= \ln M-H(X^n)+H(X^n|f)-n \epsilon\nonumber\\
%&\ge\sum_{i=1}^nR-I(Y^{i-1};Y_i)-H(X_i|X_{i+1}^n)+H(X_i|f,X^n_{i+1})- \nonumber\\
%&=\sum_{i=1}^nR-I(Y^{i-1},X_{i+1}^n;Y_i)+I(X_{i+1}^n;Y_i|Y^{i-1})-H(X_i|Y^{i-1},X_{i+1}^n)-I(Y^{i-1};X_i|X_{i+1}^n)+H(X_i|f,X^n_{i+1})- \nonumber\\
%&=\sum_{i=1}^nR-I(Y^{i-1},X_{i+1}^n;Y_i)-H(X_i|Y^{i-1},X_{i+1}^n)+H(X_i|f,X^n_{i+1})- \nonumber\\
%&\overset{\ref{c31}}{\ge}\ln M+\sum_{i=1}^n-I(Y^{i-1},X_{i+1}^n;Y_i)-H(X_i|Y^{i-1},X_{i+1}^n)+H(X_i|f,Y^{i-1},X^n_{i+1})- \epsilon\nonumber\\
%&=\ln M+\sum_{i=1}^n-I(Y^{i-1},X_{i+1}^n;Y_i)-I(X_i;f|Y^{i-1},X_{i+1}^n)-\epsilon \nonumber\\
%&=\ln M+\sum_{i=1}^n-I(U_i;Y_i)-I(X_i;Z_i|U_i)- \epsilon
\end{align}
where
\begin{enumerate}
\item Due to Fano's inequality. \label{c31}
%\item According to an argument same as (\ref{rate}). \label{c32}
\end{enumerate}
We have
\begin{align}
\ln|g|+\ln|h|
&\ge H(g,h)\nonumber\\
&\ge I(g,h;X^n,Y^n)\nonumber\\
&\ge I(X^n;g,h)+I(Y^n;g,h|X^n)\nonumber\\
&= H(X^n)-H(X^n|g,h)+I(Y^n;g,h|X^n)\nonumber\\
&\overset{\ref{c41}}{\ge} \ln M-n\epsilon +I(Y^n;g,h|X^n)\nonumber\\
&\overset{\ref{c42}}{=} \ln M-n\epsilon +I(Y^n;g|X^n)\nonumber\\
&=\ln M +\sum_{i=1}^n -\epsilon +I(Y_i;g|X^n,Y^{i-1})\nonumber\\
&\overset{\ref{c43}}{=}\ln M+\sum_{i=1}^n - \epsilon+I(Y_i;g|X_i,Y^{i-1},X_{i+1}^n)\nonumber\\
&=\ln M+\sum_{i=1}^n -\epsilon +I(Y_i;Z_i|X_i,U_i)\label{rategh}
\end{align}
\begin{enumerate}
\item Due to Fano's inequality. \label{c41}
\item $h$ is a deterministic function of $X^n$ \label{c42}
\item $g$ is a deterministic function of $Y^n$. Due to the memoryless property, we have\label{c43}
\begin{align}
H(g|X_i,Y^{i-1},X_{i+1}^n, X^{i-1})&=H(g|X_i,Y^{i-1},X_{i+1}^n)\\
H(g|Y_i, X_i,Y^{i-1},X_{i+1}^n, X^{i-1})&=H(g|Y_i, X_i,Y^{i-1},X_{i+1}^n)
\end{align}
\end{enumerate}

We note that $\frac{1}{n}\ln M\ge R-\eta$, $\frac{1}{n}\ln|g|\le R_1+\eta$ and $\frac{1}{n}\ln|h|\le R_2+\eta$, for an arbitrary $\eta>0$. Assume $ \epsilon\rightarrow 0$, then from (\ref{rate}), (\ref{rateg}), (\ref{rateh}) and (\ref{rategh}), we have
\begin{align}
R&\le \frac{1}{n}\sum_{i=1}^nI(U_i;Y_i)+H(X_i|U_i)\\
R_1&\ge\frac{1}{n}\sum_{i=1}^n I(Z_i;Y_i|U_i,X_i)\\
R_2&\ge \frac{1}{n}\sum_{i=1}H(X_i|U_i,Z_i)\\
R_1+R_2&\ge R+\frac{1}{n}\sum_{i=1}^n I(Y_i;Z_i|X_i,U_i)
\end{align}
Define a time-sharing random variable $Q$, which is uniformly distributed on $\{1,2,\dots,n\}$. Also define a set of random variables $(X,Y,\tilde{U},\tilde{Z})$ such that
\begin{align}
Pr(X=x,Y=y,\tilde{U}=u,\tilde{Z}=z|Q=i)=p(X_i=x,Y_i=y,&U_i=u,Z_i=z),\quad &i=1,2,\dots,n
\end{align}
Define $U=(\tilde{U},Q)$ and $Z=(\tilde{Z},Q)$, then
\begin{align}
R&\le \frac{1}{n}\sum_{i=1}^nI(U_i;Y_i)+H(X_i|U_i)\nonumber\\
&=I(\tilde{U};Y|Q)+H(X|\tilde{U},Q)\nonumber\\
&\le I(\tilde{U},Q;Y)+H(X|\tilde{U},Q)\nonumber\\
&=I(U;Y)+H(X|U)\label{dcpr1}
\end{align}
\begin{align}
R_1&\ge\frac{1}{n}\sum_{i=1}^n I(Z_i;Y_i|U_i,X_i)\nonumber\\
&=I(\tilde{Z};Y|\tilde{U},Q,X)\nonumber\\
&=I(Z;Y|U,X)\label{dcpr2}
\end{align}
\begin{align}
R_2&\ge\frac{1}{n}\sum_{i=1}H(X_i|U_i,Z_i)\nonumber\\
&=H(X|\tilde{U},\tilde{Z},Q)\nonumber\\
&= H(X|U,Z)\label{dcpr3}
\end{align}
\begin{align}
R_1+R_2&\ge R+\frac{1}{n}\sum_{i=1}^n I(Y_i;Z_i|X_i,U_i)\nonumber\\
&=R+I(\tilde{Z};Y|\tilde{U},X,Q)\nonumber\\
&=R+I(Z;Y|U,X)\label{dcpr4}
\end{align}
%which concludes the proof of the converse part.
where (\ref{dcpr1}), (\ref{dcpr2}), (\ref{dcpr3}) and (\ref{dcpr4}) are the same as (\ref{dc1}), (\ref{dc2}), (\ref{dc3}) and (\ref{dc4}), concluding the proof.

Finally, we note that the bounds on the cardinalities of the alphabets  in (\ref{dcb1}) and (\ref{dcb2}) can be proven in a way similar to \cite[Appendix D]{Kaspi:1982}.

\section{Remarks}
We have several remarks regarding this result as follows:
\begin{enumerate}
\item The capacity is strictly smaller than the cut-set bound \cite{Cover:1991}, because first
\begin{equation}
R\le R_1+R_2-I(Y;Z|U,X)
\end{equation}
An operational interpretation is that when the noisy relay cannot fully decode the message, or in other words, when the noisy relay cannot remove the noise completely, the data going through the link from the noisy relay to the receiver contains noise. Thus, the useful information flowing through the multiple access cut will be strictly less than $R_1+R_2$. Secondly, we note that 
\begin{equation}
R\le I(U;Y)+H(X|U)\le H(X)
\end{equation} 
An operational interpretation is that when the noisy relay decodes the message with a positive rate, the rate of information flowing through the broadcast cut becomes strictly less than $H(X)$.

Consider the following example. Let $X$ and $Y$ be binary and 
\begin{equation}
Y=X\oplus W
\end{equation}
where the sum is a modulo-$2$ sum and $W$ has a Bernoulli distribution with entropy $0.5$ bits. We assume $R_1=R_2=0.5$ bits. The cut-set bound in this example is $1$ bit, which is not achievable. Because if $R$ is equal to $1$ bit, we have,
\begin{equation}
R= I(U;Y)+H(X|U)= H(X)=1
\end{equation}
then, $U$ has to be independent of $X$ and $Y$. Also, we have 
\begin{equation}
R= R_1+R_2-I(Y;Z|U,X)=R_1+R_2=1
\end{equation}
then, $Z$ has to be independent of $X$ and $Y$ if $U$ is independent of $X$ and $Y$.
However,  if $U$ and $Z$ are independent of $X$ and $Y$, we arrive at the following contradiction,
\begin{equation}
0.5=R_2\ge H(X|Z,U)=H(X)=1
\end{equation}
which means that the cut-set bound is not achievable in this example. We note that, even in this binary example where $|\mathcal{X}|=|\mathcal{Y}|=2$, the cardinalities of the auxiliary random variables $U$ and $Z$ are $|\mathcal{U}|\le6$ and $|\mathcal{Z}|\le 15$. These large cardinality bounds make it practically impossible to evaluate the capacity of this diamond channel. However, we note that,
even though we were not able to compute the exact value of the capacity in this example, we were able to conclude that the capacity is strictly less than the cut-set bound, which is $1$ bit. 
 
We know that the capacity of a diamond channel with four orthogonal links is equal to the cut-set bound in this channel. Our result shows that introducing the broadcast node will reduce the capacity of this all-orthogonal diamond channel. 
Networks with broadcast nodes have been studied recently from different perspectives, e.g.,~information theory and network coding \cite{Dana: 2006, Ratnakar:2006, Kramer:2008}. 
We note that our diamond channel model is a simple example of a general network with a broadcast node. Thus, we conclude that the cut-set bound in general is not tight in networks with broadcast nodes.

\item The processing at the noisy relay includes two operations: decode the inner code $U^n$ and compress the channel output $Y^n$ to $Z^n$ conditioned on $U^n$. This processing is essentially the same as Theorem $7$  in \cite{Cover:1979}, i.e.,~combination of DAF and CAF. DAF \cite[Theorem 1]{Cover:1979} has been shown to be optimal in the degraded relay channel \cite{Cover:1979}. Partial DAF, a special case of  \cite[Theorem 7]{Cover:1979} without compression, has been shown to be optimal in semi-deterministic relay channel \cite{El_Gamal:1982} and the relay channel with orthogonal transmitter-relay link \cite{El_Gamal:2005}.  Recently, CAF \cite[Theorem 6]{Cover:1979} has been shown to be optimal in two special relay channels \cite{Kim:2008, Aleksic:2007}. To our knowledge, we are the first to show the optimality of the combination of DAF and CAF in some specific channel, even though the channel we consider is not a three-node relay channel in the strict sense, i.e.,~as in \cite{Cover:1979}.

\item If we assume $R=H(X)-R_0$, then Theorem \ref{maintheo} can be rewritten as follows
\begin{align}
&R\le I(U;Y)+H(X|U)&\longleftrightarrow&\qquad R_0\ge I(U;X|Y)\label{dkb1}\\
&R_1\ge I(Z;Y|U,X)&\longleftrightarrow&\qquad R_1\ge I(Z;Y|U,X)\label{dkb2}\\
&R_2\ge H(X|Z,U)&\longleftrightarrow&\qquad R_2\ge I(X;X|Z,U)\label{dkb3}\\
&R_1+R_2\ge R+I(Y;Z|X,U)&\longleftrightarrow&\qquad R_0+R_1+R_2\ge I(X,Y;U,X,Z)\label{dkb4}
\end{align}
for some joint distribution 
\begin{equation}
p(u,z,x,y)=p(u,x)p(y|x)p(z|u,y)\label{dkb5}
\end{equation}
We note that the right hand sides of (\ref{dkb1}), (\ref{dkb2}), (\ref{dkb3}) and (\ref{dkb4}) in addition to the distribution constraint in (\ref{dkb5}) are the same as the rate region of the rate-distortion problem studied by Kaspi and Berger as shown in Figure \ref{Kaspiberger} \cite[Theorem 2.1, Case C]{Kaspi:1982}.

\begin{figure}
\centering
\input{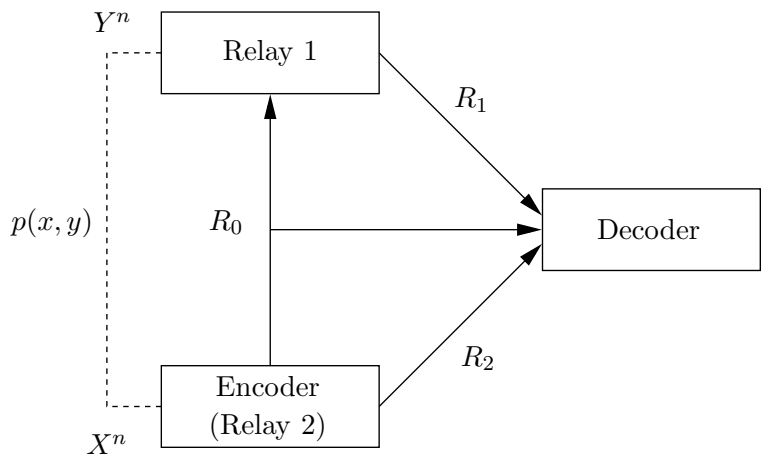}
\caption{Kaspi-Berger rate distortion problem.}
\label{Kaspiberger}
\end{figure}

This duality between our diamond channel coding problem and the Kaspi-Berger source coding problem is similar to the duality between the single-user channel coding problem and the Slepian-Wolf source coding problem\cite[Section 3.1]{Csiszar:1981}  by viewing the codebook information in the channel coding problem as the information sent to all the terminals in the source coding problem, e.g.,~the information with rate $R_0$ in Figure \ref{Kaspiberger}. Thus, the achievability of our diamond channel coding problem can be obtained from the achievability of Kaspi-Berger source coding problem, in the same way that the achievability of the multiple access channel coding problem can be obtained from the achievability of fork network coding problem \cite[Section 3.2]{Csiszar:1981}. 
\end{enumerate}

\bibliographystyle{unsrt}
\bibliography{refphd}

\begin{thebibliography}{10}

\bibitem{Schein:2001}
B.~E. Schein.
\newblock {\em Distributed Coordination in Network Information Theory}.
\newblock PhD thesis, Massachusetts Institute of Technology, 2001.

\bibitem{Csiszar:1981}
I.~Csisz\'{a}r and J.~K\"{o}rner.
\newblock {\em Information Theory: Coding Theorems for Discrete Memoryless
  Systems}.
\newblock Academic Press, 1981.

\bibitem{Csiszar:1978}
I.~Csisz\'{a}r and J.~K\"{o}rner.
\newblock Broadcast channels with confidential messages.
\newblock {\em IEEE Trans. Inform. Theory}, 24(3):339--348, 1978.

\bibitem{Kaspi:1982}
A.~H. Kaspi and T.~Berger.
\newblock Rate-distortion for correlated sources with partially separated
  encoders.
\newblock {\em IEEE Trans. Inform. Theory}, 28(6):828--840, 1982.

\bibitem{Cover:1991}
T.~M. Cover and J.~A. Thomas.
\newblock {\em Elements of Information Theory}.
\newblock John Wiley and Sons, 1991.

\bibitem{Dana:2006}
A.~F. Dana, R.~Gowaikar, R.~Palanki, B.~Hassibi, and M.~Effros.
\newblock Capacity of wireless erasure networks.
\newblock {\em IEEE Trans. Inform. Theory}, 52(3):789--804, 2006.

\bibitem{Ratnakar:2006}
N.~Ratnakar and G.~Kramer.
\newblock The multicast capacity of deterministic relay network with no
  interference.
\newblock {\em IEEE Trans. Inform. Theory}, 52(6):2425--2432, 2006.

\bibitem{Kramer:2008}
G.~Kramer, S.~M. S.~Tabatabaei Yazdi, and S.~A. Savari.
\newblock Network coding on line networks with broadcast.
\newblock In {\em Proc. Conf. Inf. Sciences and Systems (CISS)}, Princeton, NJ,
  Mar. 2008.

\bibitem{Cover:1979}
T.~M. Cover and A.~El~Gamal.
\newblock Capacity theorems for the relay channel.
\newblock {\em IEEE Trans. Inform. Theory}, 25:572--584, Sep. 1979.

\bibitem{El_Gamal:1982}
A.~El~Gamal and M.~Aref.
\newblock The capacity of the semideterministic relay channel.
\newblock {\em IEEE Trans. Inform. Theory}, 28(3):536, 1982.

\bibitem{El_Gamal:2005}
A.~El~Gamal and S.~Zahedi.
\newblock Capacity of a class of relay channels with orthogonal components.
\newblock {\em IEEE Trans. Inform. Theory}, 51(5):1815--1817, 2005.

\bibitem{Kim:2008}
Y.~H. Kim.
\newblock Capacity of a class of deterministic relay channels.
\newblock {\em IEEE Trans. Inform. Theory}, 53(3):1328--1329, 2008.

\bibitem{Aleksic:2007}
M.~Aleksic, P.~Razaghi, and W.~Yu.
\newblock Capacity of a class of modulo-sum relay channels.
\newblock {\em Submitted to IEEE Transactions on Information Theory}, 2007,
  http://arxiv.org/pdf/0704.3591.

\end{thebibliography}
\end{document}